\documentclass[twocolumn,showpacs,aps,prl,superscriptaddress]{revtex4}

\usepackage{xspace}
\usepackage{graphicx}
\usepackage{dcolumn}
\usepackage{amsmath}
\usepackage{epsfig}

\input babarsym.tex

\newcommand{\roots}        {\ensuremath{\sqrt{s}}\xspace}

\newcommand{\eetautau}   {\ensuremath{e^+e^-\to\tautau}\xspace}
\newcommand{\eemm}       {\ensuremath{e^+e^-\to\mumu}\xspace}
\newcommand{\eeqq}       {\ensuremath{e^+e^-\to\qqbar}\xspace}

\newcommand{\tautomu}   {\ensuremath{\tau^\pm \to \mu^\pm \nunub}\xspace}
\newcommand{\tautoel}   {\ensuremath{\tau^\pm \to e^\pm \nunub}\xspace}

\newcommand{\tautohnpiz}  {\ensuremath{\tau^\pm \to h^\pm (\ge 1) \pi^0 \nu}\xspace}

\newcommand{\taumg}      {\ensuremath{\mtau^{\pm} \to \mmu^{\pm} \g}\xspace}
\newcommand{\muelg}      {\ensuremath{\mmu^{\pm} \to \electron^{\pm} \g}\xspace}

\newcommand{\BRtaumg}    {\ensuremath{\BR(\taumg)}\xspace}
\newcommand{\BRmuelg}    {\ensuremath{\BR(\muelg)}\xspace}

\newcommand{\ntaupair}         {\ensuremath{2.07\times 10^8}\xspace}

\newcommand{\lumion}             {\ensuremath{210.6 \invfb}\xspace}
\newcommand{\lumioff}             {\ensuremath{21.6 \invfb}\xspace}

\newcommand{\sigback}            {\ensuremath{6.2\pm 0.5}\xspace}

\newcommand{\EventsFromFit}            {\ensuremath{-2.2^{+3.2}_{-2.4}}\xspace}
\newcommand{\EventsFromFitNoError}            {\ensuremath{-2.2}\xspace}
\newcommand{\BackgroundFromFit}            {\ensuremath{143\pm 12}\xspace}
\newcommand{\BackgroundFromFitNoError}            {\ensuremath{143}\xspace}
\newcommand{\BranchingRatio}          {\BRtaumg$=(-5.6^{+8.3}_{-6.3})\times10^{-8}$\xspace}

\newcommand{\UpperLimit}            {\BRtaumg$<6.8\times10^{-8}$\xspace}
\newcommand{\UpperLimitBayes}       {\BRtaumg$<14\times10^{-8}$\xspace}

\newcommand{\datatomcegsb}       {\ensuremath{1.052\pm0.056(stat)\pm0.024(norm)}\xspace}
\newcommand{\Emg}             {\ensuremath{E_{\mu\gamma}}\xspace}
\newcommand{\Mnu}             {\ensuremath{m_{\nu}^2}\xspace}

\newcommand{\trkhel}           {\ensuremath{\cos\theta_H}\xspace}

\newcommand{\mugam}            {\ensuremath{\mu\gamma}\xspace}

\def\kk2f       {\mbox{\tt KK2f}\xspace}
\def\tauola     {\mbox{\tt Tauola}\xspace}

\newcommand{\gevccgevcc}{\ensuremath{{\mathrm{\,Ge\kern -0.1em V^2\!/}c^4}}\xspace}
\newcommand{\eff} {\ensuremath{\varepsilon}}

\newcommand{\evcc}{\ensuremath{{\mathrm{\,e\kern -0.1em V\!/}c^2}}\xspace}
\newcommand{\CM} {\mbox{c.m.}\xspace}

\newcommand{\BABARPubYear}     {04}
\newcommand{\BABARPubNumber}  {049}

\newcommand{\SLACPubNumber} {11028}
\newcommand{\LANLNumber}  {0502032}

\def\figurebox#1#2#3{%
    \def\arg{#3}%
    \ifx\arg\empty
    {\hfill\vbox{\hsize#2\hrule\hbox to #2{\vrule\hfill\vbox to #1{\hsize#2\vfill}\vrule}\hrule}\hfill}%
    \else
    {\hfill\epsfbox{#3}\hfill}%
    \fi}

\begin{document}

\preprint{\babar-PUB-\BABARPubYear/\BABARPubNumber} 
\preprint{SLAC-PUB-\SLACPubNumber} 

%% Needed in final document
\begin{flushleft}
% For internal BABAR ANALYSIS DOCUMENT
%\begin{tabular}{lcr}
%\babar-PUB-\BABARPubYear/\BABARPubNumber & ~~~~~~~~~~~~~~~~~~~~~~~~~~~~~~~~~~~~~~~~~~~~~~~~~~~~~~~
%   & \mbox{\normalsize {\babar\ }Analysis Document \#1079, Version 9}
%\end{tabular}\\
\babar-PUB-\BABARPubYear/\BABARPubNumber\\
SLAC-PUB-\SLACPubNumber\\
hep-ex/\LANLNumber\\[10mm]    % Comment out until a hep-ex number is assigned
\end{flushleft}

\title{
%% Needed only for BAD drafts
%\begin{flushleft}
%       \mbox{\normalsize {\babar\ }Analysis Document \#1079, Version 10}
%       \end{flushleft}
%       \vskip 20pt
{\large \bf \boldmath
Search for Lepton Flavor Violation in the Decay \taumg} 
}

% author list; 
%% author list as of 02-Dec-2004 (617 authors)
%
\author{B.~Aubert}
\author{R.~Barate}
\author{D.~Boutigny}
\author{F.~Couderc}
\author{Y.~Karyotakis}
\author{J.~P.~Lees}
\author{V.~Poireau}
\author{V.~Tisserand}
\author{A.~Zghiche}
\affiliation{Laboratoire de Physique des Particules, F-74941 Annecy-le-Vieux, France }
\author{E.~Grauges-Pous}
\affiliation{IFAE, Universitat Autonoma de Barcelona, E-08193 Bellaterra, Barcelona, Spain }
% ...
\collaboration{The \babar\ Collaboration}
\noaffiliation

\date{\today}% It is always \today, today, but you may specify any date with \date.

\begin{abstract}
A search for the nonconservation of lepton flavor number
 in the decay \taumg has been performed using \ntaupair \eetautau
events produced at a center-of-mass energy near
 10.58\gev\ with the \babar\ detector at the PEP-II storage ring.
We find no evidence for a signal and 
set an upper limit on the branching ratio of \BRtaumg\ $<6.8\times10^{-8}$ at 90\% confidence level.
\end{abstract}


\maketitle

%%
%% --------- Introduction ----------------
%%

 Decays violating lepton flavor number, if observed, would be
 among the most theoretically clean signatures of new physics and
 the decay \taumg is one such  process. It is expected with rates as high as
 several parts per million in some supersymmetric models \cite{Hisano:1998fj,King:1998nv}, 
 despite the stringent experimental limit on  the related \muelg decay \cite{Brooks:1999pu}. 
 In a modest extension to the Standard Model (SM) incorporating
 finite $\nu$ masses \cite{NuOsc},
 the branching ratio is many orders of magnitude below experimental accessibility \cite{Lee:1977ti}, 
 and so an observation of this mode would unambigously indicate new physics.
 Currently the most stringent limit is \BRtaumg$<3.1\times10^{-7}$ at 90\% confidence level (c.l.)
 from the BELLE experiment \cite{Abe:2003sx}.

%%
%% --------- Data set and Detector ----------------
%%
The search for \taumg decays reported here uses data recorded by the \babar\ detector at the SLAC \pep2
 asymmetric-energy \epem storage ring.
 The data sample consists of an integrated luminosity of \L = \lumion\ recorded at a
 center-of-mass energy (\roots) of $\sqrt{s}=10.58\gev$,
 and  \lumioff\ recorded at  $\sqrt{s}=10.54\gev$.
The luminosity-weighted average cross section for \eetautau is
$\sigma_{\tau\tau}=(0.89\pm0.02)$ nb \cite{kk},
corresponding to a data sample of \ntaupair $\tau$-pair events.

The \babar\ detector is described in detail in Ref. \cite{detector}.
Charged particles are reconstructed as tracks with 
a 5-layer silicon vertex tracker and a 40-layer drift chamber (DCH)
inside a 1.5-T solenoidal magnet.
An electromagnetic calorimeter (EMC) consisting of 6580 CsI(Tl) 
crystals is used to identify electrons and photons.
The flux return of the solenoid, instrumented with resistive 
plate chambers (IFR), is used to identify muons.

%%
%% --------- Selection ----------------
%%
The signature of the signal process is the presence of an isolated \mmu and \g 
having an invariant mass consistent with that of the \mtau (1.777\gevcc \cite{bes}) and a total
energy (\Emg) equal to \roots/2 in the event center-of-mass (\CM) frame, 
and properties of the other particles in the event which are consistent with a SM \mtau decay.
Such events are simulated with higher-order radiative corrections using the \kk2f Monte Carlo (MC) generator
 \cite{kk} where one $\tau$ decays into \mugam according to phase space \cite{flatphasespace},
 while the other $\tau$ decays
according to measured rates \cite{PDG} simulated with \tauola \cite{tauola,photos}.
The detector response is simulated with \mbox{\tt GEANT4} \cite{geant}. 
The simulated events for signal as well as SM background
processes \cite{kk,tauola,photos,Lange:2001uf,Sjostrand:1995iq}
are then reconstructed in the same manner as data.
The MC backgrounds are used for selection optimization and efficiency systematic studies,
but not for the final background estimation, which relies solely on data.

Events with two or four well reconstructed tracks and zero net charge are selected.
The magnitude of the thrust vector calculated with all observed charged
and neutral particles, characterising the direction of maximum energy
flow in the event~\cite{thrust}, is required to lie 
between 0.900 and 0.975 to suppress \eeqq backgrounds with low thrust and \eemm and Bhabha backgrounds with
thrust close to unity.
Other non-$\tau$ backgrounds are suppressed by requiring 
the polar angle ($\theta_{miss}$) of the missing momentum associated with the neutrinos in the event
to lie within the detector acceptance $(-0.76 < \cos \theta_{miss} < 0.92)$,
and the scaled missing \CM transverse momentum relative to
the beam axis $(p^{T}_{miss}/\sqrt{s})$ to be greater than 0.068 (0.009)
for events with two (four) tracks.

The signal-side hemisphere, defined with respect to the thrust axis, is required to
contain one track with  \CM momentum less than 4.5\gevc\  and
 at least one \g\ with a \CM energy greater than 200\mev . 
The track must be identified  as a \mmu using DCH, EMC and IFR information 
and the \g\ candidate is the one which gives the mass of the \mugam system closest to
 the \mtau mass. This provides the correct pairing for $99.9\%$ of selected signal events.
The resolution of the \mugam mass is improved  by assigning 
the point of closest approach of the \mmu track to the \epem collision axis
as the origin of the \g candidate and by using a kinematic fit with 
\Emg\ constrained to \roots/2. This energy-constrained mass (\mec) and
 $\DeltaE =\Emg -\roots/2$ are independent variables 
apart from small correlations arising from initial and final state radiation.
The mean and standard deviation of the \mec and \DeltaE distributions for reconstructed 
MC signal events are: $\langle \mec \rangle$ = 1777\mevcc, $\sigma(\mec)$ = 9\mevcc, 
$\langle \DeltaE \rangle$ = $-$9\mev, $\sigma(\DeltaE)$ = 45\mev, 
where the shift in $\langle \DeltaE \rangle$ comes from photon energy reconstruction effects.
We blind the data events within a $3\sigma$ ellipse centered on
$\langle \mec \rangle$ and $\langle \DeltaE \rangle$ 
until completing all optimization and systematic studies of the selection criteria.

The dominant backgrounds are from \eemm  and \eetautau (with a \tautomu\ decay)
 events with an energetic $\gamma$
from initial or final state radiation or in the $\tau$ decay.
For these backgrounds, the $\gamma$ is predominantly along the $\mu$ flight direction;
thus we require $|\trkhel| < 0.8$, where $\theta_H$ is the angle 
between the \mmu momentum in the reconstructed $\tau$ rest frame and the $\tau$ momentum in the laboratory frame.
Backgrounds arising from \tautohnpiz decays with the hadronic track $(h)$ mis-identified as a \mmu,
are reduced by requiring the total \CM energy of non-signal $\gamma$ candidates
in the signal-side hemisphere to be less than 200\mev.
If the reconstructed neutral candidate identified as the signal $\gamma$, has at least 1\% likelihood of arising from overlapping daughters in $\pi^0 \rightarrow \g\g$ decays, then the event is rejected.

The tag-side hemisphere, which is expected to contain a SM \mtau decay, is required 
to have a total invariant mass  less than 1.6\gevcc 
and a \CM momentum for each track less than 4.0\gevc 
to reduce background from \eeqq and \eemm processes, respectively.
The  $\qqbar$ background is further reduced by requiring the hemisphere to have
no more than six $\gamma$ candidates.

 A tag-side hemisphere containing a single track is
 classified as e-tag,  $\mu$-tag or  h-tag 
if the total photon \CM energy in the hemisphere is no more
 than 200\mev and the track is exclusively identified as an electron (e-tag), as a muon ($\mu$-tag)
 or as neither (h-tag). If the total photon \CM energy in the hemisphere is more
 than 200\mev, then events are selected if the track is exclusively identified as an electron
 (e$\gamma$-tag) or as neither an electron nor as a muon (h$\gamma$-tag).
 These allow for the presence of radiation in \tautoel decays and for photons from 
 $\pi^0 \rightarrow \g\g$ in \tautohnpiz decays.
 If the tag-side contains three tracks, the event is classified as a 3h-tag.
 We explored other tag-side channels but the sensitivity of the search does not improve
 by including them.

Hadronic $\tau$ decays have only one missing $\nu$, a feature used to
purify the sample. Taking the tag-side $\tau$ direction to be opposite the 
fitted signal $\mugam$ candidate, we use all tracks and $\gamma$ candidates 
on the tag-side to calculate the invariant mass squared of the missing $\nu$ ($\Mnu$),
and require $|\Mnu|$  to be less than 0.4\gevccgevcc for h-tag and 3h-tag events and
less than 0.8\gevccgevcc for h$\gamma$-tag events.

\begin{figure}
\vspace*{3mm}
\resizebox{\columnwidth}{.52\textheight}{
\includegraphics{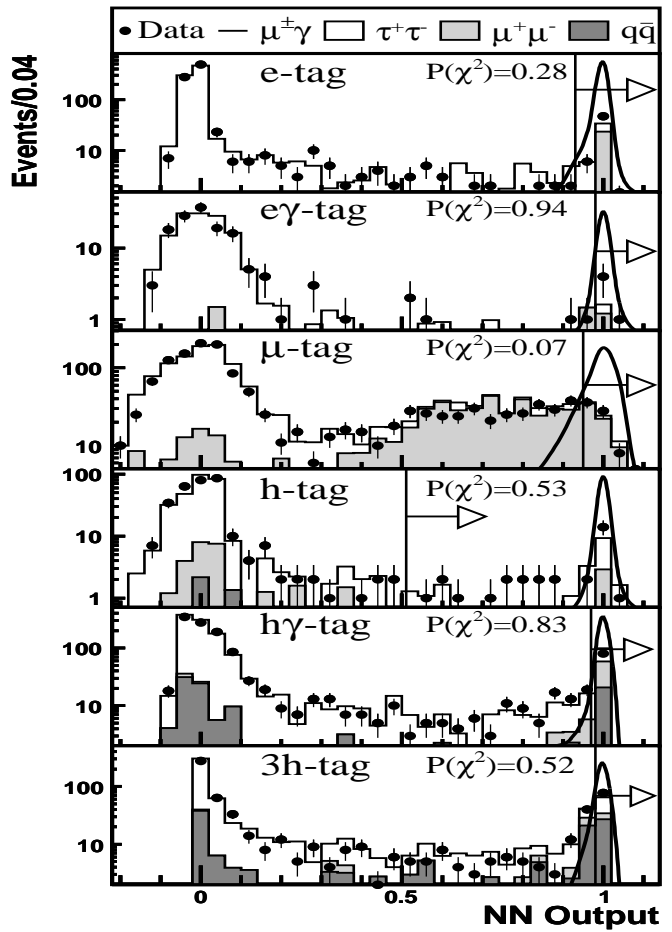}}
\caption{NN output shown for data (dots),
MC backgrounds (histograms normalized to the luminosity) and MC signal 
(curves with arbitrary normalization) in the GSB region.
Lines with arrows indicate optimized cut positions.
The probability of the data-MC $\chi^2$ is indicated for each tag mode.}
\label{fig1}
\end{figure}

At this stage of the analysis, 15\% of the MC signal events survive within 
a Grand Side Band (GSB) region defined as: 
\mec $\in$ [1.5, 2.1] \gevcc, \DeltaE $\in$ [-1.0, 0.5] \gev.
The non-blinded part of the GSB contains 4688 data events,
which agrees with the MC background expectation of 4924 events to within 5\%.
Out of these MC events, 80\% are from \eetautau,
 82\% of which are \tautomu decays on the signal-side.

To further suppress the backgrounds,
separate neural net (NN) based discriminators are used for each of the six tags.
Five observables serve as input to the NN:
the missing mass of the event,
the highest \CM momentum of the tag-side track(s), \trkhel, $p^T_{miss}$ and \Mnu.
Each NN is trained using data in the non-blinded part of the GSB to describe the background 
and $\mu\gamma$ MC in the full GSB region to describe the signal.
The NN output distributions of the data (Figure~\ref{fig1}) are in good agreement
 with MC backgrounds both in shape and absolute rates, as are the input observables.
The MC signal within a $2\sigma$ ellipse in the \mec-\DeltaE plane centered
on $\langle \mec \rangle$ and $\langle \DeltaE \rangle$, 
and the MC background interpolated from \mec sidebands 
($|\mec - \langle \mec \rangle|> 3 \sigma$ within the GSB and 
$|\DeltaE - \langle \DeltaE \rangle| < 3 \sigma$) 
are then used to optimize the cut value on the NN output based on the expected 90\%~c.l. upper limit.
The optimized NN cut values are restricted to be $>0.5$.  
Within the $\pm 3\sigma$ band in \DeltaE, the MC predicts that 66\% of the selected background comes
from \eemm, 27\% from \eetautau and the rest from  $\epem \ra \qqbar$ processes.

\begin{figure}
 \resizebox{.92\columnwidth}{.3\textheight}{%  was .25
\includegraphics{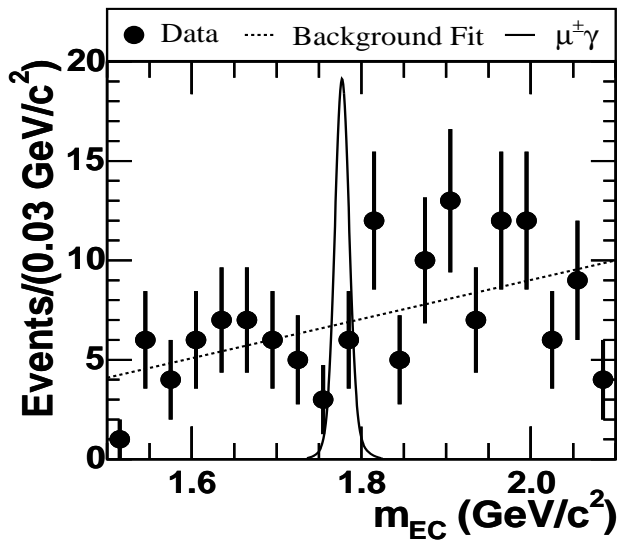}}
\caption{\mec distribution of data (dots),
 the background component of the fit (dotted line) and
 MC signal  (curve with arbitrary normalization) 
 for $|\DeltaE - \langle \DeltaE \rangle| < 2 \sigma$.
The $\chi^2$ between data and the background component is 16.0 for 20 bins.}
\label{fig2}
\end{figure}

With the data unblinded, we find four events in the $2\sigma$ signal ellipse
where we expect \sigback events, obtained from a linear interpolation of the data in the \mec sidebands.
Other polynomials up to at least fifth order predict the same level of background
to within half a standard deviation. 
The agreement between observed data and background expectations 
across the different tagging modes are shown in  Table~\ref{tab:table1}.

The relative systematic uncertainties on the trigger efficiency,
 tracking and photon reconstruction efficiencies, and particle identification 
are estimated to be 1.2\%, 1.3\%, 1.8\% and 1.2\%, respectively.
We obtain a measure of the systematic error of the efficiency due to
simulation uncertainties of the NN input variables 
by fixing each input variable to its average value one at a time, 
without retraining or changing the architecture of the NN, and re-calculating the efficiency. 
This has the effect of removing each input variable completely from the NN selection 
and gives a 1.9\% relative error on the signal efficiency.
Adding these errors in quadrature gives 3.4\%.
As we use $1.2\times10^6$ MC signal events, the contribution to the
 error arising from signal MC statistics is negligible.

\begin{table}
\begin{center}
\begin{tabular}{l|l|r|r|r|r|r|r|r} \hline
\multicolumn{2}{r|}{Tag:}
          & e     &e$\gamma$&$\mu$& h    &h$\gamma$& 3h &  all  \\\hline\hline
   &  Data     & 57    & 6     & 67    & 31   & 92    & 78   & 331  \\
GSB&  MC      & 46.4  & 2.7   & 63.1  & 19.2 & 108.9 & 64.9 & 305\\ \cline{2-9}
   &  \eff (\%) & 1.88  & 0.27  & 1.80  & 1.44 & 3.72  & 1.85 & 11.0 \\ \hline\hline

2$\sigma$    & Selected     & 1     & 0     & 1     & 0    & 1     & 1    & 4   \\ \cline{3-9}
signal       &  Expected     & 1.1   & 0.1   & 1.9   & 0.5  & 1.8   & 0.9  & 6.2 \\
ellipse  & from Data       &$\pm$0.2 &$\pm$0.1 &$\pm$0.3 &$\pm$0.1 &$\pm$0.3&$\pm$0.2&$\pm$0.5 \\
   \cline{2-9}
          & \eff (\%) & 1.27  & 0.18  & 1.31  & 0.89 & 2.56  & 1.22 & 7.42 \\ \hline\hline

$\pm2\sigma$         & Data     & 20    & 0     & 51    & 9    & 41    & 20   & 141  \\ 
   \cline{2-9}
in \DeltaE           & \eff (\%) & 1.62  & 0.23  & 1.63  & 1.13 & 3.22  & 1.53 & 9.35  \\ \hline\hline

\end{tabular} 
\caption{Number of events for data and MC backgrounds  
for the different tags in the full GSB;
in the $2\sigma$ signal ellipse, the number of
events selected in data  and expected from the data sidebands; 
the number of data events selected  inside the $\pm 2\sigma$ band in \DeltaE; 
and the respective efficiencies (\eff).}
\label{tab:table1}
\end{center}
\end{table}

Alternatively, these (and other potential sources of systematic uncertainty not necessarily
accounted for in the above procedure) can be collectively estimated
from the detector modelling uncertainty obtained by comparing  data to the MC backgrounds
in the non-blinded part of the GSB, where the background and signal have similar properties
apart from \mec and \DeltaE.
Data and MC background statistics  as well as signal  efficiencies (\eff) 
are shown in Table~\ref{tab:table1} inside the full GSB.
The agreement in the GSB between data and the background MC for each tag-mode and their combination
validates the ability of the MC to simulate these signal-like events.
The statistical precision of this data-to-MC ratio is augmented by using the expanded range \mec $\in [1.0, 2.5] \gevcc$ 
to obtain a value of \datatomcegsb in the non-blinded part of the GSB. 
To be conservative, we quote the total 6.1\% uncertainty on this ratio,
which includes a 2.3\% normalization error on the product $\L\sigma_{\tau\tau}$,
as the relative systematic error on the \eff\ in the GSB.%without the \mec and \DeltaE requirements.

To obtain the branching ratio, 
we  perform an extended unbinned maximum likelihood (EML) fit to the \mec data distribution 
(Figure~\ref{fig2}) after all requirements but that on \mec have been applied. Within this
 $\pm 2\sigma$ band in \DeltaE the efficiencies for the different tag-modes are
given in  Table~\ref{tab:table1} for a total value of
 \eff~=~(9.4$\pm$0.6)\%, where the systematic error here includes
 an additional contribution from the \DeltaE requirement.
A linear parameterization describes the background and a double Gaussian serves as the
probability density function (PDF) of the signal.
 Uncertainities in the mean and resolution of \mec are incorporated
 into the fit by convoluting the signal PDF with another
 Gaussian with $\sigma=4\mevcc$
and by increasing the $\sigma$ of the convoluted Gaussian by 1\mevcc. 
The quoted limit is insensitive to these variations, however.

In the EML fit, the number of signal events is given by 
2$\L\sigma_{\tau\tau}$\eff\BRtaumg and we fit for the branching ratio, the number of background
events and slope of the background. The systematic uncertainty on \eff\ is incorporated
into the likelihood by adding \eff\ as a fourth fit parameter  under the constraint that it
follows a Gaussian spread about its measured value within the estimated errors.
This yields the same upper limit as the fit without
the constraint on \eff\ to within the quoted number of significant figures.
The fit gives \BranchingRatio, 
which corresponds to \EventsFromFit signal and \BackgroundFromFit background events.
From the likelihood function of this fit, a Bayesian upper limit 
can be derived \cite{Bayes}.

In keeping with established \taumg studies \cite{Abe:2003sx,Ahmed:1999gh},
we derive a frequentist upper limit \cite{Narsky:1999kt}. We generate MC samples with
Poisson-distributed numbers of signal and background events. The 
expected number of background events is fixed to \BackgroundFromFitNoError and we scan over
the expected number of signal events, $s$. The \mec values 
are distributed according to the signal and background PDFs, where the background
slope is generated from a Gaussian distribution with mean and standard deviation
given by the fit to the data. The number of signal events in each sample is extracted 
using the same EML fit procedure as that applied to the data. We vary $s$ until we find a
value for which 90\% of the sample yields a fitted number of  signal events greater
than that observed in the data, {\it i.e.} \EventsFromFitNoError.
At 90\% c.l. this procedure  gives an upper limit of \UpperLimit 
\cite{FrequencyofLimit}.

As confirmation of this result, we also undertake an analysis without the NN, having
 the same sensitivity of $12\times10^{-8}$ for the expected 90\% c.l. upper limit.
Events with a tag-side muon are vetoed but single-track tag events are
 otherwise not classified.
Cuts are applied on the signal $\mu$ momentum, signal $\gamma$ energy, 
$\theta_{miss}$, $p^T_{miss}$, the tag-side
invariant mass and \DeltaE, and  \mec is required to be within 30\mev\ of $m_{\tau}$.
This selection retains  10.7\% of the signal and has a background of 28.5$\pm$2.3 events  
 as  estimated from the sidebands.

To enhance the signal/background discrimination,
a likelihood-ratio variable, $\mathcal R$, is built from
four discriminating variables: $p^T_{miss}$, \DeltaE, the difference
between the signal $\mu$ and $\gamma$ energy in the c.m.,
 and the acoplanarity between the signal \mugam system and the tag system.
We observe no evidence of signal and we compare the 
 two-dimensional  (\mec, $\mathcal R$) distribution
of the 27 events  in data with the background and signal expectations,
 utilizing  a classical frequentist  $\mathrm {CL}_{S+B}$ method~\cite{junk}. 
The limit set is consistent with the above value and
    amounts to a 90\% c.l. limit of $9.4\times 10^{-8}$.

% Input the pubboard acknowledgements file
We are grateful for the excellent luminosity and machine conditions
provided by our \pep2\ colleagues, 
and for the substantial dedicated effort from
the computing organizations that support \babar.
The collaborating institutions wish to thank 
SLAC for its support and kind hospitality. 
This work is supported by
DOE
and NSF (USA),
NSERC (Canada),
IHEP (China),
CEA and
CNRS-IN2P3
(France),
BMBF and DFG
(Germany),
INFN (Italy),
FOM (The Netherlands),
NFR (Norway),
MIST (Russia), and
PPARC (United Kingdom). 
Individuals have received support from the 
A.~P.~Sloan Foundation, 
Research Corporation,
and Alexander von Humboldt Foundation.


\begin{thebibliography}{99}
\bibitem{flatphasespace}
Although the signal MC has been modelled using a phase space model, the limit obtained
in this analysis is insensitive to this assumption as demonstrated by considering the two 
extreme cases of a V-A and a V+A form of interaction for the signal MC.
The insensitivity is largely a consequence of the symmetric cut on $|\trkhel|$.
\bibitem{Bayes}
A Bayesian upper limit assuming a uniform prior can be obtained by
integrating the likelihood function of the fit, $L$, from zero to the value of the 
branching ratio that includes 90\% of  $\int_0^{\infty}L(\BR)d\BR$.
This gives a 90\% c.l. upper limit of  \UpperLimitBayes.
Integrating from $-\infty$ intead of zero changes this to 7.9$\times 10^{-8}$.
\bibitem{FrequencyofLimit}
For an equivalent experiment with no signal,
the probability of obtaining a 90\% c.l. upper limit at or below $6.8\times10^{-8}$
is 30\%.
\end{thebibliography}
\end{document}